\documentclass[a4paper]{mem}
\usepackage{natbib}
\usepackage{graphicx}
\usepackage[a4paper]{hyperref}
\idline{1}{1}
\def\ltsim{\hbox{\raise 2pt \hbox {$<$} \kern-1.1em \lower 4pt \hbox {$\sim$}}}
\def\ltapprox{\hbox{\raise 2pt \hbox {$<$} \kern-1.1em \lower 5pt \hbox 
{$\approx$}}}
\def\gtsim{\hbox{\raise 2pt \hbox {$>$} \kern-1.1em \lower 4pt \hbox {$\sim$}}}
\def\gtapprox{\hbox{\raise 2pt \hbox {$>$} \kern-1.1em \lower 5pt \hbox 
{$\approx$}}}

\def\arcsec{$^{\prime\prime}$}
\def\arcmin{$^{\prime}$}
\def\degrees{$^{\circ}$}
\begin{document}
   \title{Properties of Cluster Radio Emission
}

   \author{Luigina Feretti 
}
   \institute{Istituto di Radioastronomia CNR,
via P. Gobetti 101, 40129 Bologna, Italy \email{lferetti@ira.cnr.it}\\ 
             }

   \abstract{
Relevant studies of the non-thermal components of the intracluster
medium are  performed at radio wavelengths.
A number of clusters, indeed, exhibits cluster-wide diffuse radio emission,
which is indication of the existence of large scale magnetic fields and
of relativistic electrons in the cluster volume. 
There is strong evidence that the presence of diffuse radio emission is 
related to cluster merger processes.
The details of the halo-merger connection are discussed and 
a brief outline of current models of halo formation is presented.
   }
   \authorrunning{L. Feretti}
   \titlerunning{Properties of Cluster Radio Emission}
   \maketitle
%

\section{Introduction}

   \begin{figure*}
   \centering
   \includegraphics[width=16cm]{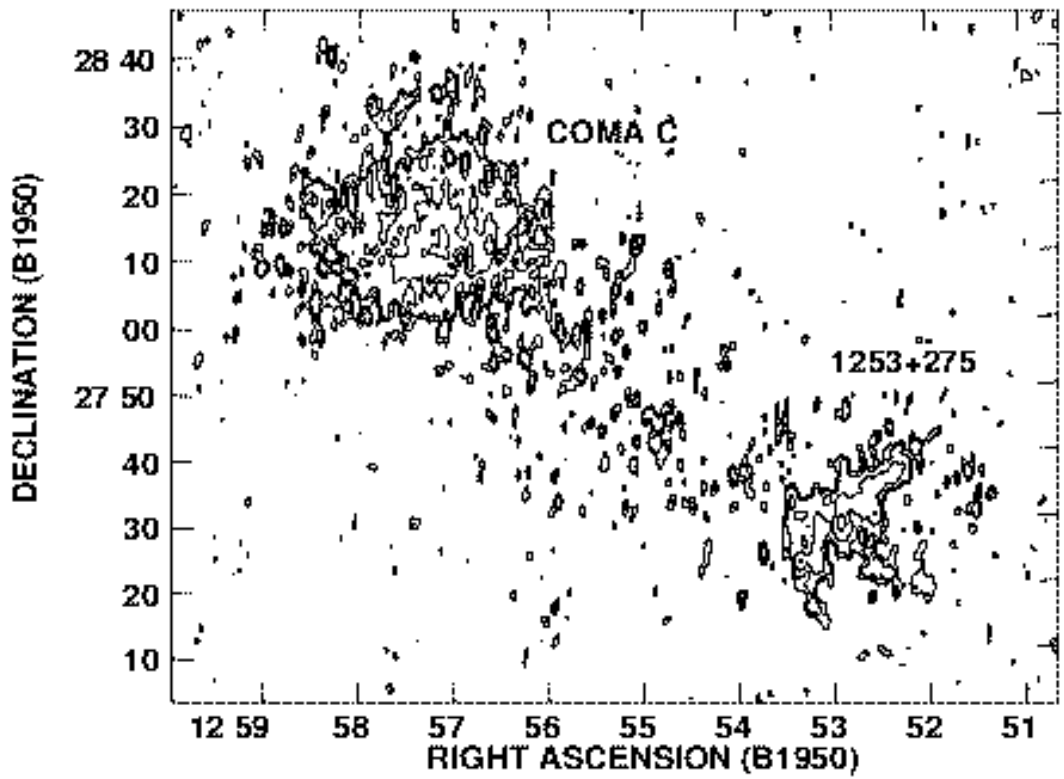}
   \caption{Diffuse radio emission in the 
Coma cluster, obtained at 90 cm with the Westerbork 
Synthesis Radio Telescope. The discrete sources have been subtracted.
The cluster center is approximately located at the
position RA$_{1950}$ = 12$^h$ 57$^m$ 24$^s$, DEC$_{1950}$ = 28\degrees
15\arcmin 00\arcsec. The radio halo Coma C is at the
cluster center, the radio relic 1253+275 is at the cluster
periphery. An angular size of 10\arcmin~ corresponds to a linear size of
$\sim$ 400 kpc. Contour levels are at 2.5, 4, 8, 16 mJy/beam (FWHM =
55\arcsec $\times$ 125\arcsec; RA $\times$ DEC). 
   }
              \label{Fig1}%
    \end{figure*}

The main components of clusters of galaxies are the galaxies (2-3\%),
the hot gas (13-15\%) and the dark matter (82-85\%).  In addition, a
relativistic component may be present which 
 plays an important role in the
cluster formation and evolution.  The most detailed studies of this
component come from the radio observations.  A number of clusters of
galaxies is known to contain large-scale diffuse radio sources which
have no obvious connection with the cluster galaxies, but are rather
associated with the intracluster medium (ICM).  These sources are
classified in two groups, radio halos and relics, according to their
location at the cluster center or cluster periphery, respectively.
The synchrotron origin of the emission from these sources requires the
presence of cluster-wide magnetic fields of the order of $\sim$ 0.1-1
$\mu$G, and of a population of relativistic electrons with Lorentz
factor $\gamma >>$ 1000 and energy density of
$\sim$ 10$^{-14}$-10$^{-13}$ erg cm$^{-3}$.

The importance of halos and relics is that they are large scale
features,  related to other cluster properties in the optical
and X-ray domains, and thus  connected to the cluster
history and evolution.


These sources are found in clusters which have recently undergone a
merger event, thus leading to the idea that they originate from
particle acceleration in cluster merger turbulence and shocks.  The
formation and evolution of these sources is however still under
debate: the radio emitting electrons could be reaccelerated cosmic rays,
or accelerated from the thermal population, or could be produced as a
result of the interaction between cosmic-ray protons and the ICM.  
We summarize the current knowledge on
these sources from an observational point of view.

The instrinsic parameters quoted in this paper are computed
with a Hubble constant H$_0$ = 50 km s$^{-1}$ Mpc$^{-1}$ 
and a deceleration parameter q$_0$ = 0.5.

%
   \begin{figure}
   \centering
   \includegraphics[width=7cm]{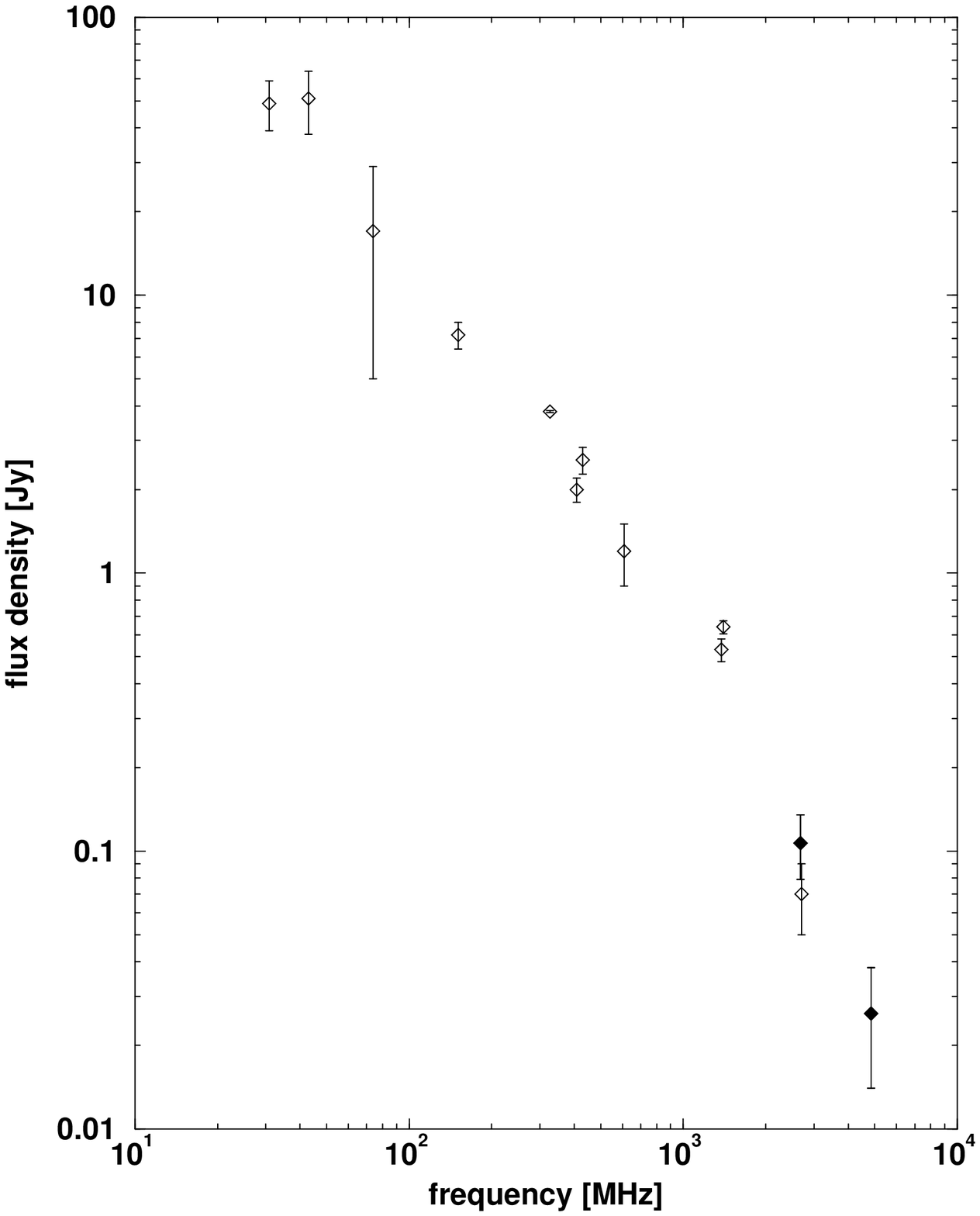}
      \caption{Total radio spectrum of the radio halo Coma C
(from Thierbach et al. 2003).
              }
         \label{Fig2}
   \end{figure}

   \begin{figure}
   \centering
   \includegraphics[width=7cm]{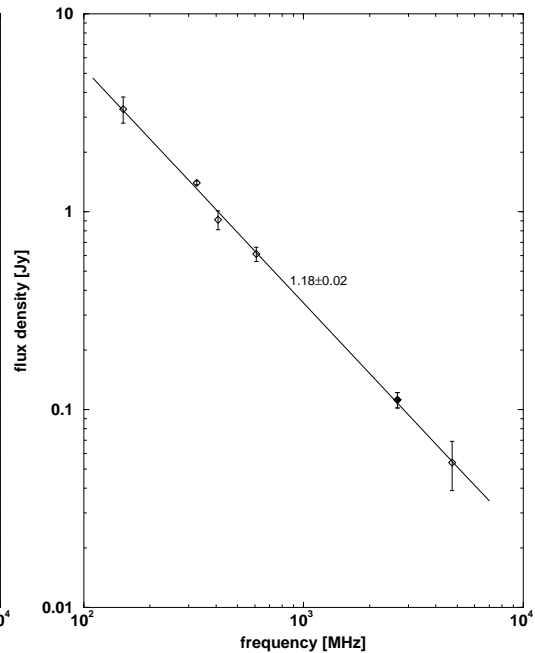}
      \caption{Total radio spectum of the relic 1253+275
(from Thierbach et al. 2003).
              }
         \label{Fig3}
   \end{figure}
%

\section {Radio halos and relics: the Coma cluster}

The Coma cluster is the first cluster where a radio halo and a relic
have been detected (Willson 1970, Ballarati et al. 1981).
The halo in this cluster, Coma C (see Fig. 1), is the prototypical
example of halo sources: it is located at the cluster center, it is
characterized by a regular shape with a total extent of $\sim$ 1 Mpc,
and by a low radio surface brightness (\ltsim~ $\mu$Jy arcsec$^{-2}$
at 1.4 GHz).  It is unpolarized down to a limit of a few percent, and
shows a steep radio spectrum, typical of aged radio sources
($\alpha$ \gtsim~ 1), with a steepening at higher frequencies
(Fig. 2).  The spectral index distribution of Coma C shows a radial
decrease (Giovannini et al. 1993) from $\alpha \sim$ 0.8 at the
cluster center, to $\alpha \sim$ 1.8 beyond a distance of about
10\arcmin.  By assuming that there is energy equipartition between
relativistic particles and magnetic field, a minimum energy density of
1.62 10$^{-14}$ erg cm$^{-3}$ is derived from the radio data.  The
corresponding equipartition magnetic field is 0.4 $\mu$G.

The radio source  1253+275 in the Coma cluster (Fig. 1)
is the prototype of the class of radio relics, which 
are extended diffuse radio sources associated with 
the ICM, located in the cluster peripheral regions.
This source is similar to the halo Coma C in its low surface brightness,
large size and steep spectrum (Fig. 3).
Unlike halos, it shows an elongated structure and it is
highly polarized ($\sim$ 25\%).

   \begin{figure*}
   \centering
   \includegraphics[bb=105 185 550 630,width=9cm,clip]{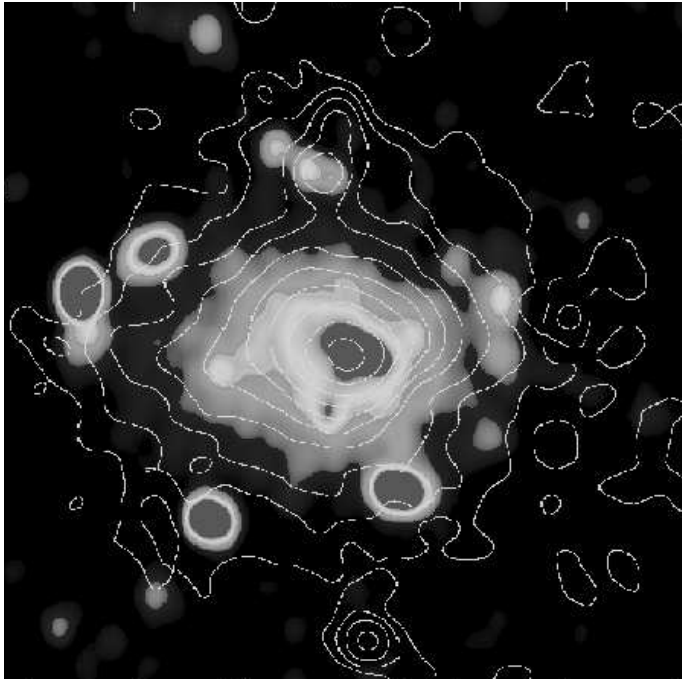}
      \caption{The cluster A2163 in radio and X-ray.  The grey
scale image represents the radio emission in A2163 at 20 cm, showing
an extended radio halo.  The contours represent the ROSAT X-ray
emission. The extended irregular X-ray structure indicates the
presence of a recent cluster merger.  
              }
         \label{Fig4}
   \end{figure*}

The radiative lifetime of the relativistic electrons in Coma C,
 considering synchrotron and inverse Compton energy losses, is
of the order of 10$^8$ yr. This is too short to allow the particle
diffusion throughout the cluster volume. This implies that the
radiating electrons cannot have been injected at some particular point
of the cluster, but they must undergo {\it in situ} energization.
This is a general problem for all the halo and relic sources.
Feretti (2002) argued that halos and relics are not the same objects
seen in projection, i.e. halos are really at the cluster center and
not simply projected onto it.  Halos and relics may indeed have
different physical origins.

Coma is one of the few clusters where hard X-ray emission has been
detected with the BeppoSAX and  Rossi X-ray Timing Explorer (RXTE)
satellites (Fusco-Femiano et al. 2004 and references therein).  
This emission is expected in clusters with diffuse radio
sources, as the high energy relativistic electrons responsible for the
radio emission ($\gamma$ $\sim$ 10$^4$) scatter
off the cosmic microwave background, boosting photons from this
radiation field to the hard X-ray domain by inverse Compton (IC) process.
Measurements of this
radiation provide additional information that, combined with
results of radio measurements (i.e. the ratio of hard X-ray IC
emission to radio synchrotron emission), enables the determination of
the electron density and mean magnetic field directly, without 
invoking equipartition.

The 20-80 keV flux in Coma is $\sim$ 1.5 10$^{-11}$ erg cm$^{-2}$
s$^{-1}$, which leads to a volume averaged intracluster magnetic field
of $\sim$ 0.2 $\mu$G (Fusco-Femiano et al. 2004).  
This value is consistent with that obtained by
the radio emission assuming equipartition (see above). It is
inconsistent,
however, with the value of $\sim$ 6 $\mu$G deduced by Feretti et al. (1995)  
from Faraday Rotation Measure (RM) data (see Sect. 4).

It is worth mentionining here that alternative models have been
suggested to explain  the hard X-ray tails
(e.g. non-thermal bremsstrahlung).  These models were motivated by the
discrepancy between the value of the ICM magnetic field derived by the
IC model and the value derived from RM.  However, these models may
have serious difficulties as they would require an unrealistic high
energy input (Petrosian 2001).

\section{Connection to cluster merger processes}

Unlike the  thermal X-ray emission, the presence
of diffuse radio
emission is not  common  in clusters of galaxies.
In a complete sample,
5\% of clusters have a radio halo source and 6\% have
a peripheral relic
source (Giovannini \& Feretti 2002).  
The detection rate of diffuse radio sources increases with
the cluster X-ray luminosity, reaching $\sim$ 35\% in clusters with
X-ray luminosity larger than $\sim$ 10$^{45}$ erg s$^{-1}$.

The optical and X-ray observations indicate that halo and relic
clusters contain strong evidence of dynamical evolution (Giovannini \&
Feretti 2002).  An example is represented by the cluster A2163, which 
hosts a powerful radio halo (Feretti et al. 2001), shown in Fig. 4.  The
connection between the formation of halos and relics and the presence
of recent/ongoing mergers in the clusters is consistent with the
relative rarity of diffuse sources, and the lack of diffuse sources in
clusters showing a massive cooling flow. Cluster mergers are among the
most energetic phenomena in the Universe, releasing gravitational
binding energies of about 10$^{64}$ erg. They generate shocks, bulk
flows, turbulence in the ICM. These processes would provide energy to
reaccelerate the radiating particles all around the cluster.

The most recent investigations of the details of the connection
between radio halos and cluster mergers are presented in the
following.

\subsection{Correlation between radio and X-ray brightness}

The comparison between radio and X-ray images of clusters reveals that
there is often a close similarity on large scale between radio
and X-ray structures. Govoni et al. (2001) first performed a
quantitative point-to-point analysis of the halo radio brightness $
F_{\rm radio}$ and the cluster X-ray brightness $F_{\rm Xray}$. They 
obtained a power law relation of the type: $ F_{\rm radio} = a F_{\rm
Xray}^b$, with  values of $b$  between 0.6 and 1.
An example of the above relation is given in Fig. 5.
Since the structure of the X-ray emission is
generally related to a cluster merger process, a close connection
between the structure of the halo  and that of the X-ray gas
supports a connection between the halo radio emission and the merger.

   \begin{figure}
   \centering
   \includegraphics[width=7cm]{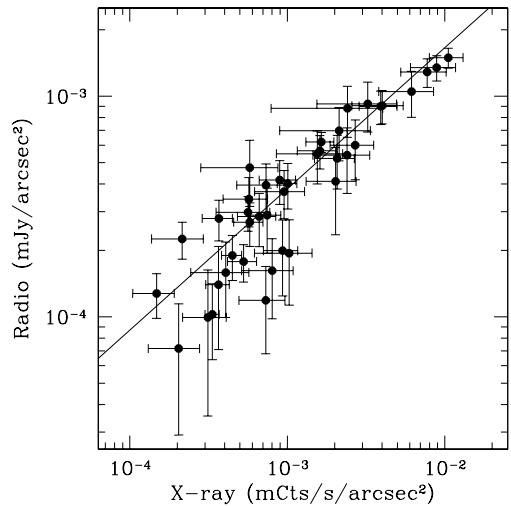}
      \caption{Plot of the radio brightness versus the X-ray brightness,
for the radio halo in A2163 (shown in Fig. 4). 
Each point represents the brightness mean in cells
of 90\arcsec~ in size, while the error bars indicate the rms in each 
cell. The best fit, indicated by the solid line, corresponds to the 
relation F$_{\rm radio}$ $\propto$ F$_{\rm Xray}^{0.64}$ (from 
Feretti et al. 2001).}
         \label{Fig5}
   \end{figure}

\subsection{Comparison between radio emission and gas temperature}

High resolution Chandra X-ray data have been recently obtained for
several clusters with halos or relics.  In all these clusters,
temperature gradients and gas shocks are detected confirming the
presence of mergers (Markevitch \& Vikhlinin 2001, Markevitch et
al. 2003, Govoni et al. 2004).  In some clusters there is a
correlation between the radio halo emission and the hot gas regions
(Govoni et al. 2004). Although it may be difficult to disentangle the
geometry of the cluster merger, it can be generally concluded that
merger shocks and turbulence are the relevant acceleration mechanisms for
the halo generation.

\subsection{Connection between radio spectrum and X-ray features}

Maps of the radio spectral index represent a powerful tool
to study the properties of the relativistic electrons and of the
magnetic field in radio sources. 
Preliminary maps of the radio spectral index between 0.3 GHz
and 1.4 GHz of the radio halos in  A665 and A2163 
(Feretti et al.  in preparation)
show a different behaviour across the regions presently interested 
by the ongoing merger and the relatively undisturbed regions.
The synchrotron spectrum is flatter in the merging regions, indicating
that the radio emitting electrons are more energetic here.
In the more relaxed regions, the spectral index steepens
progressively with the distance from the cluster center.
This global behaviour indicates that the energy of radio halos
is sensitive to the effects of mergers.

   \begin{figure}
   \centering
   \includegraphics[width=7cm]{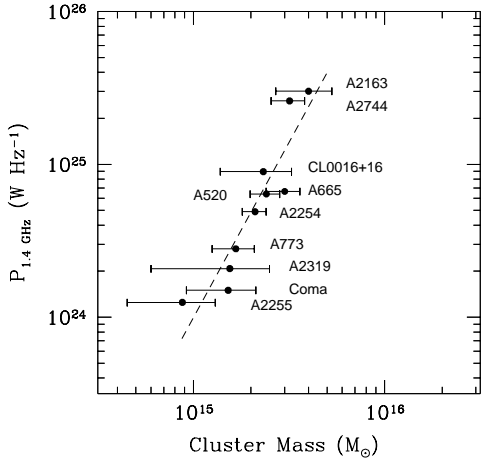}
      \caption{ Halo monochromatic radio power at 
1.4 GHz versus  total gravitational mass 
within 3 Mpc radius.
All the data are from  ROSAT, except A2163 from 
XMM (Arnaud, priv. comm.).
The best fit line (dashed) is P$_{1.4 GHz}$ $\propto$ M$^{2.3}$.
}
         \label{Fig6}
   \end{figure}

\subsection{Connection between radio power and cluster mass}

The energy available from a merger is approximately proportional to the
squared cluster mass M$^2$.  From simple arguments, it is deduced that 
the energy released in a merger shock is proportional to the gas
density $\rho$ and to the subcluster velocity $v^3$. Since $\rho
\propto$ M, and $v \propto {\rm M}^{1/2}$, one obtains that roughly
\.E $\propto$ M$^{5/2}$ (see also Kempner \& Sarazin 2001).  Fig. 6
shows that a correlation exists 
between the halo monochromatic radio power P at 1.4 GHz and the total 
gravitational cluster mass. The best fit is P$_{1.4
GHz}$ $\propto$ M$^{2.3}$. This is similar to what expected from
the above simple considerations, thus favouring the hypothesis
that the radio halo is powered by the energy released in the
cluster merger.

\section{How common are magnetic fields in clusters}

Besides the direct evidence obtained from the
synchrotron emission, the existence of cluster magnetic fields
can be indirectly probed by Rotation Measure studies of
radio galaxies embedded within the cluster thermal atmospheres
or located behind them.

This kind of studies has been performed on several individual
clusters, with or without cooling flows, e.g.  Coma, A119, A514,
A2255, 3C129, 3C295, Hydra A, as well as on statistical samples (see
the review by Carilli and Taylor 2002, and references therein).  In
general, the suggestion from the data is that magnetic fields in the
range of 1-5 $\mu$G are common in clusters, regardless of the presence
of diffuse radio emission.  At the center of cooling flow
clusters, magnetic field strenghts can be larger, up to 10-30 $\mu$G.

Another result of the above mentioned studies is that the RM
distributions tend to be patchy with coherence lenghts of 2-10 kpc,
indicating that the magnetic fields are not ordered on cluster
scales, but consist of cells with random field orientation.

The magnetic field intensity in clusters shows a radial decline.
This has been deduced in  Coma (Brunetti et al. 2001)
and in A119 (Dolag et al. 2001). In the latter cluster,
the magnetic field intensity scales with the gas density $n_e$ 
as $n_e^{0.9}$. 

The magnetic field strengths obtained from RM studies
are about an order of magnitude higher than the equipartition 
values obtained  from the radio data, or those derived
from the hard X-ray IC emission (Sect. 2).

However, the  different  measurements of the magnetic fields 
could reflect the value of the field in different regions of the clusters.
For example, since the magnetic field strength has a radial decrease,
most of the IC emission will come from the weak field
regions in the outer parts of the cluster, while most of the Faraday
rotation and synchrotron emission
occur in the strong field regions in the inner parts of the
cluster. Moreover, the magnetic field inferred from RM data
could be affected by local magnetic field compression and enhancements
(Rudnick \& Blundell 2003). 

Recent modeling by Govoni \& Murgia (2004)
shows that the presence of 
magnetic field substructure and/or filamentation
can lead to significant differences between field estimates obtained
from different approaches. Thus the above mentioned discrepancies
can be at least partially solved if a more realistic magnetic field
geometry is taken into account.

\section{Models for Relativistic Particles}

A population of relativistic electrons can account for the radio
emission in radio halos and the hard X-ray emission in clusters via
synchrotron and inverse Compton processes, respectively. Current
models have been reviewed by Brunetti (2003). The relativistic
particles could be injected in the cluster volume from AGN activity
(quasars, radio galaxies, etc.), or from star formation in normal
galaxies (supernovae, galactic winds, etc).  Most of the particle
production has occurred in the past and is therefore connected to the
dynamical history of the clusters.
This population of {\it primary electrons} needs to be reaccelerated
(Brunetti et al. 2001, Petrosian 2001) to compensate the radiative
losses. A recent cluster merger is the most
likely process acting in the reacceleration of relativistic particles.

Another class of models for the radiating particles in halos involves
{\it secondary electrons}, resulting from inelastic nuclear collisions
between the relativistic protons and the thermal ions of the ambient
intracluster medium.  The protons diffuse on large scale because
their energy losses are negligible.  They can continuously produce in
situ electrons, distributed through the cluster
volume (Blasi \& Colafrancesco 1999, Miniati et al. 2001).

A strong observational evidence in favour of
primary electron models is the behaviour of radio spectra in halos.
The high frequency steepening observed in Coma (Thierbach et al. 2003) and
in A754 (Bacchi et al. 2003, Fusco-Femiano et al. 2003), and the radial
steepening observed in Coma (Giovannini
et al. 1993), A665 and A2163 (Sect. 3.3) can be easily reproduced by
models invoking reacceleration of particles. On the contrary, the spectral
index trends are difficult to explain by models considering secondary electron
populations. Other arguments favouring primary electron
models are the observed link between radio halos and cluster mergers, 
the slope of the correlation between radio and X-ray brightness
(see Govoni et al. 2001 for details), and
 the relatively low number of clusters with halos.
On the other hand, strong $\gamma$-ray emission, which should
be detected by future $\gamma$-ray instruments, is expected to be 
produced in the framework of secondary electron models,
thus providing an important  test for the origin of the electrons
radiating in radio halos. 

Different models have been suggested for the origin of the
relativistic electrons radiating in the radio relics, i.e.  located in
confined peripheral regions of the clusters.  There is increasing
evidence that the relics are tracers of shock waves in merger
events (En{\ss}lin et al. 1998), confirming that the cluster
merger is the most important ingredient for the formation of 
any type of diffuse radio emission.

\section{Conclusions}

The existence of cluster-wide diffuse radio emission indicates 
that there are important non-thermal components in the ICM:
magnetic fields and  relativistic particles.
The presence of magnetic fields in galaxy clusters is additionally
demonstrated by RM studies, which indicate that magnetic fields are
rather common in all clusters, not only those with radio halos.
There is convincing evidence that radio halos and relics are
linked to cluster merger processes.
Violent mergers provide the energy necessary to reaccelerate
the radio emitting electrons.

\begin{acknowledgements}
I thank the workshop organizers for their invitation to this
interesting and enjoyable meeting. This work has been partially
funded by the Italian Space Agency (ASI).
\end{acknowledgements}

\bibliographystyle{aa}

\end{document}